\documentclass[lettersize,journal]{IEEEtran}
\usepackage{amsmath,amsfonts}

\usepackage{algorithm}

\usepackage{array}
\usepackage[caption=false,font=normalsize,labelfont=sf,textfont=sf]{subfig}
\usepackage{textcomp}
\usepackage{stfloats}
\usepackage{url}
\usepackage{verbatim}
\usepackage{graphicx}
\usepackage{cite}
\def\BibTeX{{\rm B\kern-.05em{\sc i\kern-.025em b}\kern-.08em
    T\kern-.1667em\lower.7ex\hbox{E}\kern-.125emX}}
\usepackage{balance}

\usepackage{multirow}
\usepackage{mathtools}
\usepackage{algorithmicx}
\usepackage{algpseudocode}
\usepackage{stfloats}

\usepackage[figuresright]{rotating}

\usepackage{booktabs}
\usepackage{makecell}
\usepackage{diagbox}
\usepackage{orcidlink}
\usepackage[acronym,toc]{glossaries}
\newacronym{pmu-drl}{PMU-DRL}{Power Measurement Utility Deep Reinforcement Learning}
\newacronym{pmu}{PMU}{Power Measurement Utility}
\newacronym{ram}{RAM}{Random Access Memory}
\newacronym{mcu}{MCU}{Microcontroller Unit}
\newacronym{njtx2}{NJTX2}{NVIDIA Jetson TX2}
\newacronym{sll}{SLL}{Sundance Lynsyn Lite}
\newacronym{pc}{PC}{Program Counter}
\newacronym{iot}{IoT}{Internet of Things}
\newacronym{sbc}{SBC}{Single Board Cumputer}
\newacronym{cgar}{CAGR}{Compound Annual Growth Rate}
\newacronym{cnn}{CNN}{Convolutional Neural Network}
\newacronym{dnn}{DNN}{Deep Neural Network}
\newacronym{dl}{DL}{Deep Learning}
\newacronym{ai}{AI}{Artificial Intelligence}
\newacronym{njn}{NJN}{NVIDIA Jetson Nano}
\newacronym{njt2}{NJT2}{NVIDIA Jetson TX2}
\newacronym{yolo}{YOLO}{You Only Look Once}
\newacronym{jtag}{JTAG}{Joint Test Action Group}
\newacronym{gpio}{GPIO}{General Purpose Input/Output }
\newacronym{arm}{ARM}{Application Response Measurement}
\newacronym{fpga}{FPGA}{Field Programmable Gate Array}
\newacronym{pbp}{PBP}{Performance Benchmark Program}
\newacronym{fps}{fps}{frames per second}
\newacronym{int8}{int8}{8 bit integer}
\newacronym{aws}{AWS}{Amazon Web Services}
\newacronym{cuda}{CUDA}{Compute Unified Device Architecture}
\newacronym{cudnn}{cuDNN}{CUDA Deep Neural Network}
\newacronym{som}{SOM}{System-on-Module}
\newacronym{gpu}{GPU}{Graphics Processor Units}
\newacronym{sdk}{SDK}{Software Developer Kit}
\newacronym{cpu}{CPU}{Central Processor Units}
\newacronym{aiot}{AIoT}{Artificial Intelligence of Things}
\newacronym{mpsoc}{MPSoC}{Multi-Processor system-on-Chip}
\newacronym{acap}{ACAP}{Adaptive Compute Acceleration Platform}
\newacronym{cv}{CV}{Computer Vision}
\newacronym{ml}{ML}{Machine Learning}
\newacronym{pmurl}{PMU-RL}{Power Measurement Utility Reinforcement Learning}
\newacronym{apc}{APC}{Average Power Consumption}
\newacronym{led}{LED}{Light Emitting Diode}
\newacronym{io}{IO}{Input/Output}
\newacronym{lslb}{LSLB}{LynSyn Lite Board}
\newacronym{iva}{IVA}{Intelligent Video Analysis}
\newacronym{ar}{ar}{augmented reality}
\newacronym{vr}{vr}{virtual reality}
\newacronym{cctv}{CCTV}{Closed-Circuit Television}
\newacronym{sm}{SM}{Streaming Multiprocessor}
\newacronym{ppo}{PPO}{Proximal Policy Optimization}
\newacronym{drl}{DRL}{Deep Reinforcement Learning}
\newacronym{ac}{AC}{Actor-Critic}
\newacronym{dvfs}{DVFS}{Dynamic Voltage Frequency Scaling}
\newacronym{avs}{AVS}{Adaptive Voltage Scaling}
\newacronym{simd}{SIMD}{Single Instruction, Multiple Data}
\newacronym{ddpg}{DDPG}{Deep Deterministic Policy Gradient}
\newacronym{sac}{SAC}{Soft Actor-Critic}
\newacronym{td3}{TD3}{Twin Delayed Deep Deterministic policy gradient algorithm}
\newacronym{trpo}{TRPO}{Trust Region Policy Optimization}

\usepackage{stfloats}
\usepackage{hyperref}

\begin{document}
\title{Deep Reinforcement Learning for Energy-Efficient on the Heterogeneous Computing Architecture}
\author{Zheqi Yu\orcidlink{0000-0002-4152-4965}, Chao Zhang\orcidlink{0000-0002-6037-0009}, Pedro Machado\orcidlink{0000-0003-1760-3871}, \IEEEmembership{Member, IEEE}, Adnan Zahid\orcidlink{0000-0001-5364-6057}, \IEEEmembership{Member, IEEE}, Tim Fernandez-Hart\orcidlink{0000-0002-8515-0002}, Muhammad Imran\orcidlink{0000-0003-4743-9136}, \IEEEmembership{Fellow, IEEE}, and Qammer Abbasi\orcidlink{0000-0002-7097-9969}, \IEEEmembership{Senior Member, IEEE}
\thanks{Manuscript created December, 2024; The authors are with the James Watt School of Engineering, University of Glasgow, Glasgow G12 8QQ, U.K. (e-mail: muhammad.imran@glasgow.ac.uk; Qammer.Abbasi@glasgow.ac.uk)
Zheqi Yu with Opteran Ltd, Sheffield, UK. (e-mail: zheqiyu@hotmail.com;) 
Chao Zhang with LAN-XEN, Technology, INC., Shanghai, China. (e-mail: ChaoZhang.z@outlook.com;) 
Pedro Machado with the Computation Intelligence and Applications Group, Department of Computer Science, Nottingham Trent University, Nottingham, U.K. (e-mail: pedro.machado@ntu.ac.uk;) 
Adnan Zahid with the School of Engineering \& Physical Sciences, Heriot Watt University, Edinburgh, EH14 4AS, U.K. (e-mail: a.zahid@hw.ac.uk;) 
Tim Fernandez-Hart with Brunel University, CEDPS, Department of Electronic and Computer Engineering, UB8 3PH, Uxbridge, UK. (e-mail:Tim.Fernandez-Hart@brunel.ac.uk)}}

\markboth{Journal of \LaTeX\ Class Files,~Vol.~xx, No.~x, December~2024}%
{Deep Reinforcement Learning for Energy-Efficient on the Heterogeneous Computing Architecture}

\maketitle

\begin{abstract}
The growing demand for optimal and low-power energy consumption paradigms for \gls*{iot} devices has garnered significant attention due to their cost-effectiveness, simplicity, and intelligibility. In this article, an \gls*{ai} hardware energy-efficient framework to achieve optimal energy savings in heterogeneous computing through appropriate power consumption management is proposed. The deep reinforcement learning framework is employed, utilising the Actor-Critic architecture to provide a simple and precise method for power saving. The results of the study demonstrate the proposed approach's suitability for different hardware configurations, achieving notable energy consumption control while adhering to strict performance requirements. The evaluation of the proposed power-saving framework shows that it is more stable, and has achieved more than 34.6\% efficiency improvement, outperforming other methods by more than 16\%.
\end{abstract}


\section{Introduction}
\label{sec:introduction}
\IEEEPARstart{T}{raditional} power management techniques rely on pre-trained \gls*{ai} algorithms and specialised energy-saving settings for different hardware. Such traditional techniques typically follow predefined energy-saving rules, such as those used in \gls*{dvfs} \cite{zhang2018double}, where the algorithm is based on feedback control energy-saving actions to optimise voltage and frequency by modelling the power and performance of the system. This approach is limited in its application and effectiveness, as it strongly depends on the construction and preprocessing of the data model. An important area of research is focused on developing methods to maximise power usage while ensuring performance without the need for predefined energy-saving methods and data pre-processing.

In our previous research \cite{yu2020energy}, Yu \emph{et al}. utilised a reinforcement learning algorithm \cite{kumar2020conservative} to observe the states of different hardware modules, and subsequently take actions to achieve energy savings. The algorithm's agent improves continuously through a specified learning method, ultimately learning the targeted optimal decisions for energy-saving actions. To enhance the performance of this method, Yu \emph{et al}. have now combined \gls*{dl} \cite{yu2022radar} with reinforcement learning, making use of the ability of neural networks to fit data \cite{9878152} and the decision-making capability of reinforcement learning \cite{wang2022deep}. The results shown a more robust deep reinforcement learning algorithm, which performs better in energy-saving tasks. Yu \emph{et al}. build upon our prior research by seamlessly integrating \gls*{dl} into an existing reinforcement learning approach. The fusion of \gls*{dl} with reinforcement learning not only enhances the performance of our energy-saving method but also leverages the data-fitting capabilities of neural networks and the decision-making prowess of reinforcement learning. The result is a more robust deep reinforcement learning algorithm that excels in energy-saving tasks. Additionally, we introduce the \gls*{pmu-drl} model for controlling hardware power consumption, marking a significant advancement from our earlier work.

In today's computational environments, heterogeneous computing systems, which combine various types of processors (such as \glspl*{cpu}, \gls*{gpu}, etc.) for enhanced performance and efficiency, have become the norm. However, managing the energy consumption of these systems presents a significant challenge. \gls*{dvfs} technology is widely regarded as an effective method to enhance the energy efficiency of heterogeneous computing hardware nodes and embedded systems. By dynamically adjusting the operating voltage and frequency of each processor, \gls*{dvfs} optimises energy consumption based on actual performance demands.

Wang \emph{et al.} \cite{wang2022energy} investigate energy savings in emerging \gls*{cpu}-\gls*{gpu} hybrid clusters through \gls*{dvfs}. They derive a rapid and precise analytical model to calculate the appropriate voltage and frequency settings for each task and employ heuristic scheduling algorithms to allocate multiple tasks to the cluster. The model underscores the nonlinear relationship between task execution time and processor speed in \gls*{gpu}-accelerated applications to capture real-world \gls*{gpu} energy consumption more accurately.

Nabavinejad \emph{et al} \cite{nabavinejad2022coordinated} introduce a fast and lightweight runtime system: BatchDVFS, that dynamically adjusts input batch processing to adaptively change batch sizes, thereby balancing throughput with power consumption. Utilising a binary search-based method to find the optimal batch size quickly, this system combines dynamic batching with \gls*{dvfs} technology to control power consumption over a broader range, thus achieving higher throughput under power constraints.

Zamani \emph{et al} \cite{zamani2023greenmd} designed an energy-saving framework: GreenMD, for an LU decomposition heterogeneous system utilising multiple \glspl*{gpu}. The predicted redundancy applies \gls*{dvfs} to heterogeneous systems by intelligently leveraging slack on \glspl*{cpu} and multiple \glspl*{gpu}. Accurate performance models for \gls*{cpu} and \gls*{gpu} are developed based on algorithmic knowledge and manufacturer’s specifications to predict slack time, achieving improved energy consumption for LU decomposition on heterogeneous multi-\gls*{gpu} systems.

However, the implementation of \gls*{dvfs} also faces challenges, including complex control logic and potential performance loss. These case studies also highlight potential drawbacks, such as the complexity of implementing \gls*{dvfs} and the need for careful tuning to balance power consumption with performance requirements. Complex control mechanisms are required to dynamically adjust voltage and frequency based on workload, which could introduce additional overheads, potentially impacting the system's responsiveness and performance.

In this article, we propose a new approach that can achieve more efficient energy management without the need for predefined energy-saving rules and data preprocessing. \gls*{drl}, with its self-learning and decision-making capabilities, provides a potential solution. \gls*{drl} can learn and improve its actions through constant observation and interaction, leading to more efficient energy management.

Wu \emph{et al.} proposes a power-saving framework: a \gls*{pmu-drl} \cite{wulf2020low} model for controlling hardware power consumption. The proposed \gls*{pmu-drl} power-saving framework is implemented on a \gls*{lslb} \cite{wulf2022rtos}, and is connected to the heterogeneous computing platform of \gls{njt2} \cite{wilson2021embedded} for implementation and evaluation. The results clearly demonstrate that excellent energy consumption control is achieved while meeting stringent performance requirements. The main contributions of this work are as follows:
\begin{itemize}
	\item Application of \gls*{drl} for Energy Optimisation: Proposing a novel energy optimisation framework that utilises \gls*{drl} to achieve significant energy savings in heterogeneous computing. This is accomplished by leveraging the data-fitting capabilities of deep learning and the decision-making proficiency of reinforcement learning, resulting in an enhanced deep reinforcement learning algorithm for superior performance in energy-saving tasks.
	\item Novel Power Management Technique: Employing a pioneering \gls*{pmu-drl} model for controlling hardware power consumption. In contrast to conventional power management techniques like \gls*{dvfs}, which rely on predefined energy-saving rules, the \gls*{pmu-drl} model eliminates the need for data pre-processing. It utilises self-adaptive feedback through the Actor-Critic architecture, simplifying and refining power-saving strategies.
	\item Empirical Evaluation: Implementing and evaluating the \gls*{pmu-drl} framework on the heterogeneous computing platform of \gls*{njtx2} . The framework exhibited enhanced stability and achieved over a 34.6\% improvement in efficiency, surpassing the performance of alternative methods.
	\item Adaptive Energy Optimisation: Developing a \gls*{drl} framework, an updated iteration of a previously employed reinforcement learning algorithm. It's self-balancing of module working states delivers an optimal energy-saving scheme for the entire system. The approach diverges from traditional handcrafted states, such as using a Q-table for reinforcement learning, and yields more substantial energy-saving effects.
\end{itemize}

The proposed method study not only offers a new energy optimisation framework, but also that this framework exhibits higher efficiency and stability, leading to significant energy savings in real-world heterogeneous computing environments.

The article's structure is organised as follows: In Section \ref{sec:methods}, the power-saving framework's operation on the heterogeneous computing platform is elucidated, along with details on the signal processing and algorithm calculation workflow. Section \ref{sec:eval} presents a quantitative evaluation of the \gls*{drl} algorithm, showcasing the achieved power-saving effects. Section \ref{sec:discussion} engages in a discussion of the power-saving results, drawing comparisons with relevant research to highlight contributions. Lastly, Section \ref{sec:conclusion} delivers a conclusion and suggests potential avenues for future research.


\section{Materials and Methods} \label{sec:methods}

The \gls*{pmu-drl} framework represents a noteworthy enhancement over the reinforcement learning algorithm applied in the prior study. Avoiding the need for data pre-processing, it utilises self-adaptive feedback via the Actor-Critic architecture. This simplification of power-saving strategies enhances precision. The framework's self-balancing of module working states delivers an optimal energy-saving scheme for the entire system, deviating significantly from traditional handcrafted states in reinforcement learning and yielding more substantial energy-saving effects.

In the proposed \gls*{pmu-drl} framework, we innovatively balance the trade-off between energy and computational performance by using power consumption information to control the working state of the \gls*{gpu} \cite{elster2022nvidia}. The proposed algorithm uniquely assesses and predicts the optimal \gls*{gpu} operational states by analysing the overall topological structure and estimating the optimal working state using power consumption data.

\begin{figure}[htbp]
\centering
\includegraphics[width=0.49\textwidth]{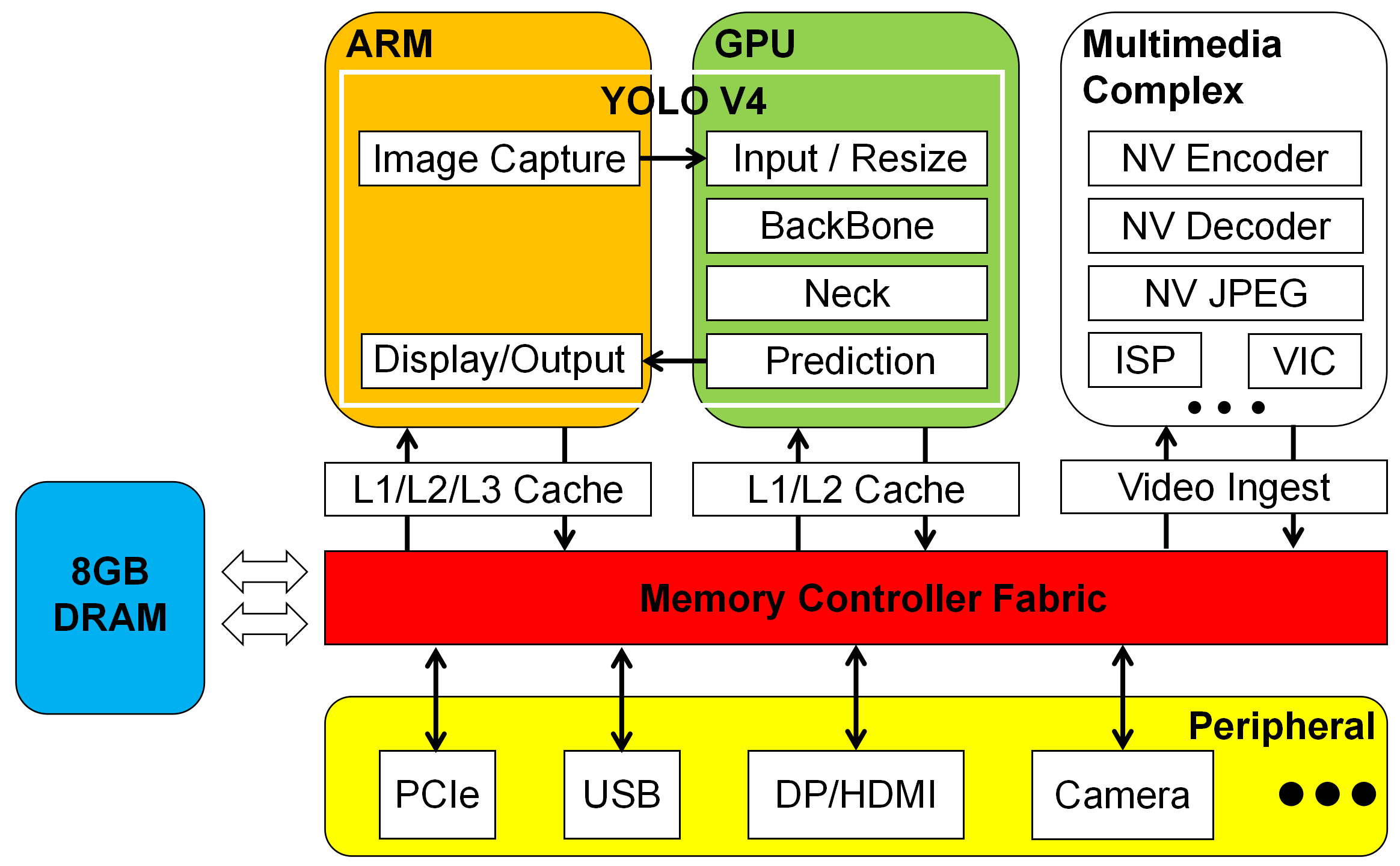}
\caption{Distributed computation stages of the \gls*{yolo}v4 neural network inference on the \gls*{njtx2}}
\label{fig:Computation_Stages}
\end{figure}

The \gls*{pmu-drl} algorithm balances the trade-off between energy and computational performance by using power consumption information to make decisions and control the working state of the \gls*{gpu}. The algorithm assesses and predicts the optimal \gls*{gpu} operational states by analysing the overall topological structure and estimating the optimal working state using power consumption data. We evaluate the performance of the \gls*{pmu-drl} by running \gls*{yolo}v4 \cite{bochkovskiy2020yolov4} on the \gls*{njtx2}, measuring the power consumption of the on-chip \gls*{cpu} and \gls*{gpu}. \gls*{yolo} is a state-of-the-art \gls*{cnn} that is able to accurately classify objects in real-time \cite{bochkovskiy2020yolov4}. The algorithm processes the entire image using a single neural network, then divides the image into parts and forecasts bounding boxes and probabilities for each object \cite{redmon2016you}. The predicted probability weighs these bounding boxes. The technique "only looks once" at the image, since it only does one forward propagation loop through the neural network before making predictions \cite{legaspi2021detection}. Figure~\ref{fig:Computation_Stages}  has shown the demonstrated \gls*{yolo}v4 algorithm workflow on the hardware.

\subsection{PMU-DRL Framework}

\begin{figure*}[hb]
\centering
\includegraphics[width=1.0\textwidth]{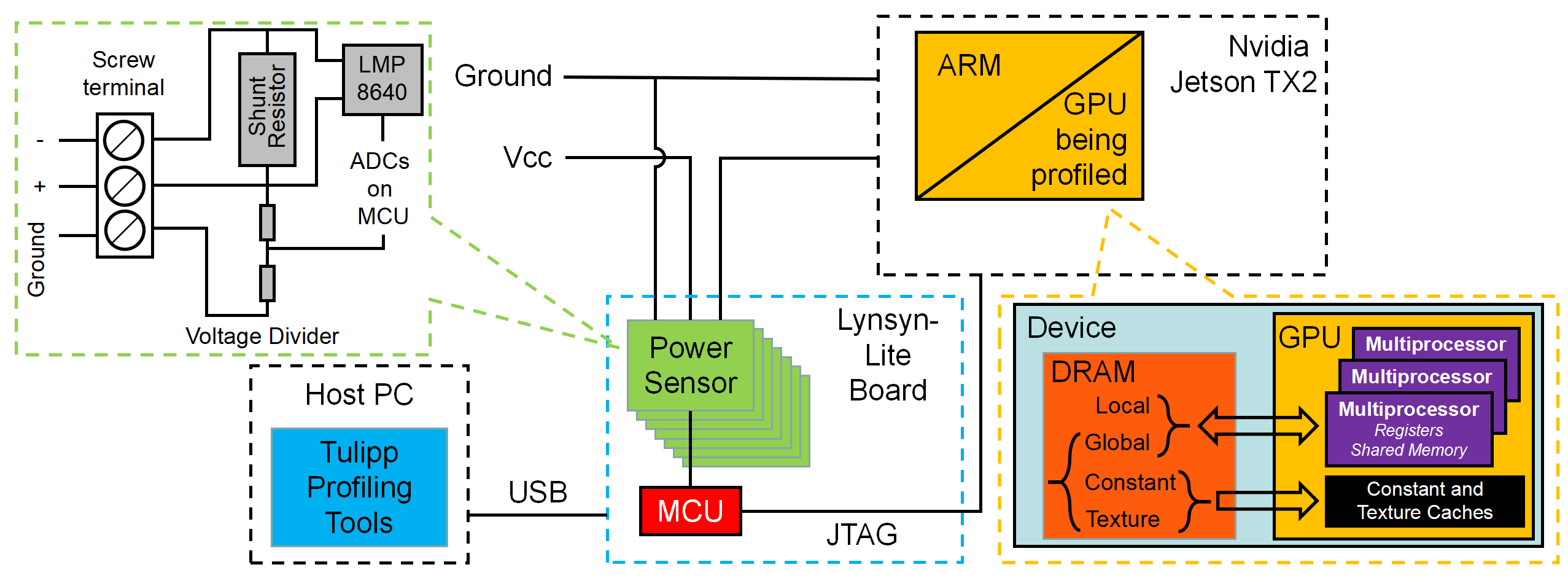}
\caption{Setup used to assess the power consumption and computational performance of \gls*{pmu-drl}. The \gls*{njtx2}  monitored by \gls*{sll}.}
\label{fig:Lynsyn_Lite}
\end{figure*}

In this section, we present the architecture of the proposed \gls*{pmu-drl} framework, which is unique in its use of a deep reinforcement learning algorithm to process power information and provide feedback for optimisation actions. Our framework makes innovative decisions to control the scheduling of the \gls*{gpu} based on the energy-saving analysis of the entire system. The setup used to evaluate the power consumption and computational performance of the \gls*{njtx2} is illustrated in Figure~\ref{fig:Lynsyn_Lite}.

Figure \ref{fig:Lynsyn_Lite} illustrates the system architecture used for evaluating the performance of the proposed \gls*{pmu-drl} framework. The setup comprises the \gls*{njtx2} as the host hardware, and a \acrfull*{sll}. for monitoring the hardware. The \gls*{sll} is equipped with a \gls*{mcu} \cite{santoni2022traveling} and power consumption sensors that are used to operate the \gls*{pmu-drl} framework. The \gls*{sll} \cite{djupdal2021lynsyn} captures voltage and current information of the \gls*{njtx2} at a rate of 1 kilo samples per second (kS/s) and sends it to the \gls*{pmu-drl} framework for analysis. The framework, based on a deep reinforcement learning algorithm, processes the power information and provides feedback for optimisation actions. Finally, the \gls*{pmu-drl} framework makes decisions to control the scheduling of the \gls*{gpu} based on the energy-saving analysis of the entire system.

In our \gls*{pmu-drl} framework, we define the following elements to formalise our problem. First, we define a state s as the current power consumption of the system, as well as the current working state of the \gls*{gpu}. Second, we define an action \textit{a} as the next operating state of the \gls*{gpu}. Finally, we define a reward function \textit{r(s, a)} as the amount of electrical energy saved by the system upon taking action a and transitioning to a new state, relative to the previous state. Our goal is to find a policy $\pi$ that selects an action \textit{a} for each state \textit{s} to maximise the expected cumulative reward. A \gls*{drl} was used to find the best policy.

In the description of the \gls*{pmu-drl} framework, it is important to note that computing performance does impact the working states of the \gls*{cpu} and \gls*{gpu}. As computing task demand fluctuates, the \gls*{cpu} and \gls*{gpu} may necessitate higher frequencies during periods of increased demand to meet performance requirements, resulting in heightened power consumption. Conversely, during decreased computing task demand, the \gls*{cpu} and \gls*{gpu} can operate at lower frequencies, thereby mitigating power consumption. Consequently, our \gls*{pmu-drl} framework dynamically adjusts the states of the \gls*{cpu} and \gls*{gpu} to optimise energy usage by analysing data on power consumption and computing performance.

\subsection{Power Measurement Unit}

\begin{figure}[htbp]
\centering
\includegraphics[width=0.49\textwidth]{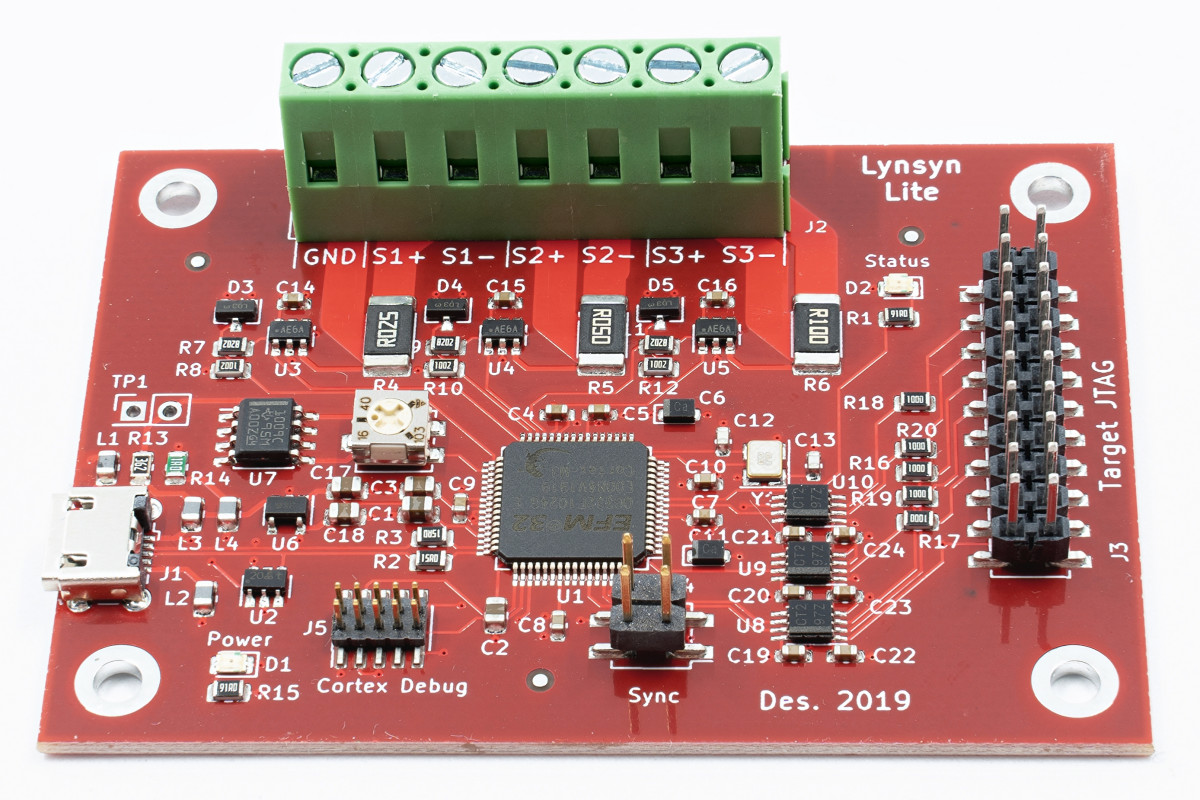}
\caption{\gls*{sll} Hardware Platform.}
\label{fig:sll}
\end{figure}

The \gls*{sll}\footnote{Available online, \protect\url{https://store.sundance.com/product/lynsyn-lite/}, last accessed 27/01/2023} (see Figure~\ref{fig:sll}) \gls*{pmu}, detailed in \cite{yu2020energy}, is purpose-built for analysing power consumption in embedded systems. While primarily optimised for \glspl*{fpga}, it can be adapted for deployment with the \gls*{njtx2}. The \gls*{sll} is equipped with three power sensors for monitoring the power consumption of the unit under test. Furthermore, it incorporates a series design featuring two USARTs (universal synchronous/asynchronous receiver/transmitter) synchronised through the EFM32 PRS trigger system \cite{callebaut2021art}. This configuration enables precise and efficient monitoring of power consumption in the \gls*{njtx2}.

In addition, when the board connects to the host workstation via \gls*{jtag}. The power measurements can be read directly from the \gls*{sll} \gls*{pc} \cite{djupdal2021lynsyn}. It can be configured to sample the target signals with frequencies up to 10kHz, and can be used as a generic remotely as an amperimeter or voltmeter. The \gls*{pmu} visualisation tool can be configured to run on a host workstation to display and control the \gls*{njtx2} via \gls*{jtag}.

The \gls*{sll} delivers efficient power measurement tools ideal to perform power analysis on embedded systems and can be used as a general power analysis tool. Therefore, the \gls*{sll} is fully compatible with major development platforms including Xilinx ZynqMP \cite{sabogal2019recon}, \gls*{njtx2} \cite{ye2023real}, Raspberry Pi 4 \cite{brunacci2023fusion}, and Sundance VCS-1 \cite{oliveira2019interactive}.

\subsubsection{Power Saving Concept for \gls*{gpu}}
Context switching is the process of switching between different tasks or threads by the processor. This is done to ensure that the maximum number of tasks are executed in the shortest possible time. When a context switch occurs, the processor saves the current task's kernel context, including the processor registers and program counter, in a Process Control Block \cite{khaing2018development} (Switch Frame). This information is then used to load the context of the new task and jump to its location in memory, allowing it to run.

The cost of context switching in a \gls*{cpu} is high, as it requires saving and restoring the data from the registers to \gls*{ram} \cite{yu2023robust}, which takes time. In contrast, the context switch of a \gls*{gpu} uses the \gls*{cpu} resources in a time-sharing manner, and is designed to process parallelisable data, which is less efficient for a \gls*{cpu} to compute sequentially. A \gls*{gpu} utilises an efficient work pool to ensure continuous task execution without idle hardware resources, and when a task encounters a pipeline stall, other tasks can run in parallel and utilise the idle pipelines. This is called a latency hiding work model, where the latency of one task is hidden by the progress of other tasks.

Context switching in a \gls*{gpu} is very fast, as it does not need to save the value of the register to memory as in a \gls*{cpu}. The \gls*{gpu} uses the \gls*{simd} mode \cite{tino2020simt}, which means that one instruction can process multiple different data, and has many cores built-in that can simultaneously process multiple threads. Additionally, the \gls*{gpu} can schedule threads at the hardware level, making context switching particularly fast with low overhead.

Overall, the \gls*{gpu} enables significantly reducing the context-switching in parallelisable tasks, which releases the \gls*{cpu} to process sequential tasks. This improves the overall performance of the system by utilising the resources of both the \gls*{cpu} and \gls*{gpu} effectively.

\subsection{Deep Reinforcement Learning}

\gls*{drl} is a type of machine learning that involves an agent learning to interact with an environment in order to achieve a specific goal. There are two main types of algorithms in \gls*{drl}, namely, 1) Behaviour policy: This is the strategy that the agent uses to interact with the environment in order to produce data; 2) Target policy: This is the strategy that the agent learns to accurately evaluate the Q value, which is the strategy that needs to be optimised.

When the two policies are the same strategy, the method is referred to as an on-policy method. When the policies are different, the method is referred to as an off-policy method. In this project, the \gls*{ppo} algorithm \cite{schulman2017proximal} was used to make smart scheduling of hardware for power optimisation. \gls*{ppo} is an on-policy \gls*{drl}, and the data used in the buffer is obtained by the target policy. This means that the Replay Buffer stores the data collected by the same policy for updating the network.

\begin{figure*}[htbp]
\centering
\includegraphics[width=0.94\textwidth]{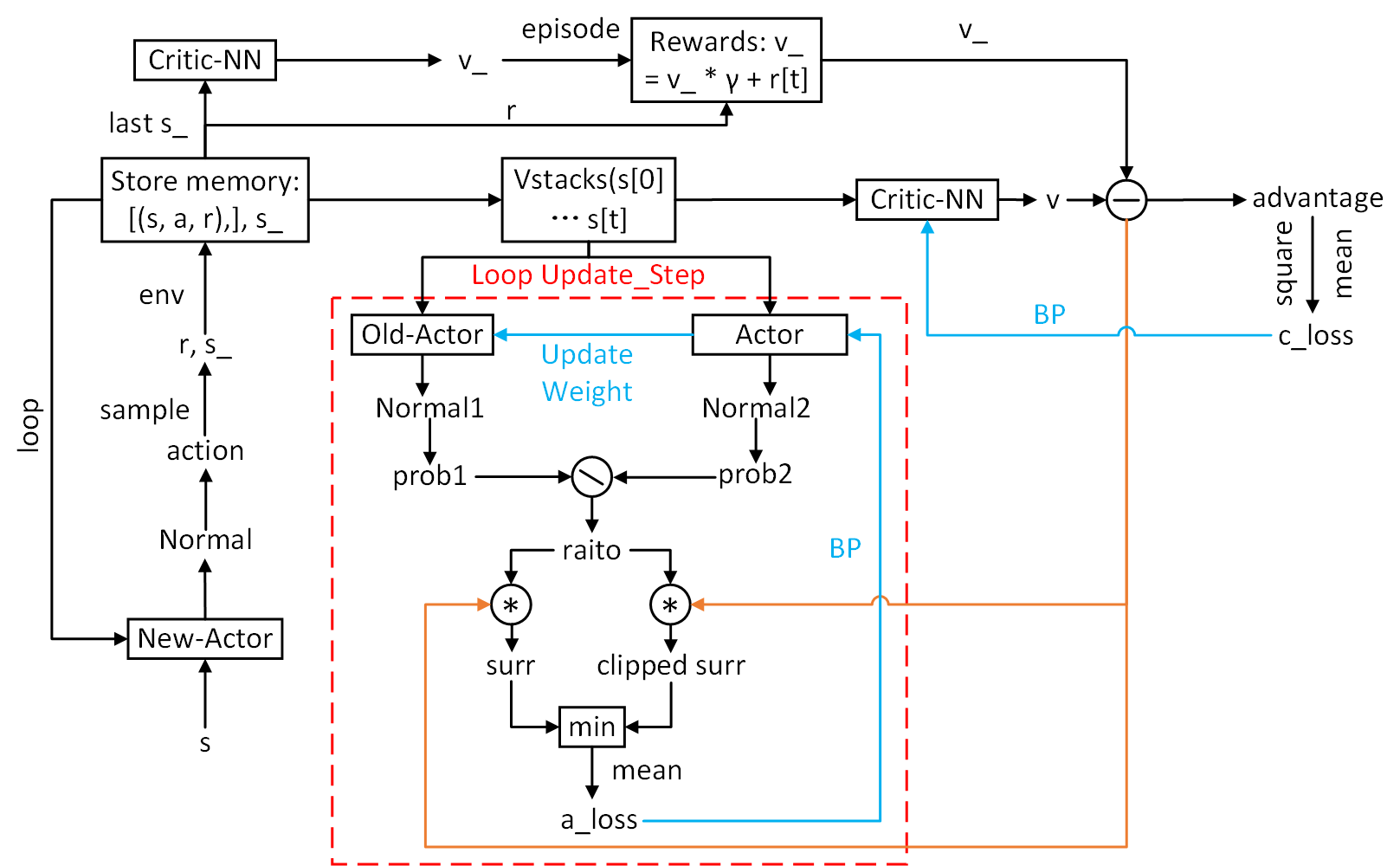}
\caption{\gls*{drl} block diagram.}
\label{fig:PPO_Structure}
\end{figure*}

The \gls*{ppo} algorithm includes two network structures: the Actor Network \cite{huang2020deep} and the Critic Network \cite{su2021value}. The Actor Network takes the state as input and outputs either the action probability for discrete action spaces or the action probability distribution parameters for continuous action spaces \cite{huang2020deep}. The Critic Network takes the state as input and outputs the state values \cite{su2021value}. The output of the algorithm is considered better if the action output by the actor network can make the advantage larger, and if the value of the state output by the critic network is more accurate.
The pseudocode is described in Algorithm~\ref{alg:Algorithm} and depicted in Figure~\ref{fig:PPO_Structure}.
The \gls*{ppo} agent consists of two parts: interaction with the environment to collect samples, and decision-making of the actions quality. One advantage of the \gls*{ac} network architecture used in \gls*{ppo} is that it can solve \gls*{drl} problems related to continuous action spaces. Unlike discrete action spaces, it does not require dictionaries or storing value functions as tables or matrices. Instead, the \gls*{ac} network architecture directly uses the Policy-Based method, which uses various policy gradient methods to directly optimise the deep neural network parameterisation's policy. This allows the deep neural network to directly output the action. Additionally, the noise variance of the action can be represented as a trainable vector, rather than being output by the network. This helps prevent overlarge noise variance from generating numerous boundary actions, which can negatively affect the performance of the agent and make the algorithm difficult to explore certain states. \gls*{ppo} is also well-suited for modelling continuous data of \gls*{gpu} energy consumption, which can be used as input for the deep reinforcement learning algorithm. The algorithm's output is the action with energy saving effect.

Figure \ref{fig:PPO_Structure} shows the deep reinforcement learning algorithm architecture. The algorithm's algorithm is illustrated in Algorithm~\ref{alg:Algorithm} and represented in Figure~\ref{fig:PPO_Structure}.

\begin{table}[htbp]
\caption{\gls*{ppo} algorithm hyper-parameters.} \centering
\label{tab:PPO_param}
\begin{tabular}{lll}
\toprule 
Hyper-parameter & \hspace{1cm} & Value \\ 
\midrule 
Sample\_Step & \hspace{1cm}  & 7000 \\
Reuse\_Times (epochs) & \hspace{1cm} & 100 \\
Gamma & \hspace{1cm} & 0.99 \\
Lambda\_Entropy & \hspace{1cm} & 0.95 \\
Clip Epsilon & \hspace{1cm} & 0.2 \\
Policy\_Learning\_Rate & \hspace{1cm} & 2e-4 \\
Value\_Function\_Learning\_Rate & \hspace{1cm} & 1e-4 \\
Train\_Policy\_Iterations & \hspace{1cm} & 32 (2048/64) \\
Train\_Value\_Iterations & \hspace{1cm} & 32 (2048/64) \\
PPO\_Update\_Times & \hspace{1cm} & 1 \\
Target\_kl & \hspace{1cm} & 2e-4 \\  
Hidden\_Sizes & \hspace{1cm} & (8, 64) \\ 
\bottomrule 
\end{tabular}
\end{table}

\begin{algorithm}[htbp]
  \caption{Proposed \gls*{pmu-drl} Algorithm}
  \label{alg:Algorithm}
  \begin{algorithmic}[1]
		\State Input \emph{s} (environmental information) into the \emph{Actor-New} network, and get two values representing the distribution of action. One is \emph{mu} (mean of Normal distribution), the other is \emph{sigma} (variance of Normal distribution).
		\State Following the Normal distribution to sample an action \emph{a}.
		\State Input \emph{s\_} into the \emph{Actor-New} network to Loop steps 1 and 2 until stored a certain amount of [(s, a, r), …]. The \emph{Actor-New} network is not updated during this process.
		\State Input the \emph{s\_} obtained in the last step of the loops into the \emph{Critic-NN} to get the \emph{v\_} value of the state. 
		\State Calculate discount rewards: R[t] = r[t] + $\gamma$ * r[t+1] + $\gamma^2$ * r[t+1] + … + $\gamma^{T-t+1}$ * r[T-1] + $\gamma^{T-t}$ * $v\_$, and then achieve R = [R[0], R[1], … ,R[t],…,R[T]]. (T is the last time step)
		\State Input all stored \emph{s} combinations into \emph{Critic-NN}, get \emph{V\_} values for all states, and then calculate $A_t$ = R – V\_
		\State c\_loss = mean(square($A_t$)), and then backpropagation updates the \emph{Critic-NN}
		\State Input all stored \emph{s} combinations into the \emph{Actor-Old} and \emph{Actor-New} networks (there are two network structure are the same), and get the Normal distribution of \emph{Normal 1} and \emph{Normal 2}.
		\State Input all stored action combinations as actions into the Normal distribution of \emph{Normal 1} and \emph{Normal 2} that achieves the corresponding probability of \emph{prob 1} and \emph{prob 2} for each action.
		\State Divide \emph{prob 2} by \emph{prob 1} to get the important weight, which is \emph{ratio}		
		\State Following equation a\_loss = mean(min((ration * $A_t$, clip(ratio, 1-$\xi$, 1+$\xi$) * $A_t$))) to update \emph{Critic-NN} by backpropagation
        \State Following some Loops by steps 8~11 to update \emph{Actor-Old} network from \emph{Actor-New} network’s weight.
		\State Looping all above steps to achieve the finally output.
  \end{algorithmic}
\end{algorithm}

The parameters for \gls*{drl} are detailed in Table \ref{tab:PPO_param}. Leveraging power consumption information from the environment, the algorithm tailors scheduling strategies for both \gls*{cpu} and \gls*{gpu}. In the environment, we define a continuous action space that represents the control strategy for power allocation to the \gls*{cpu} and \gls*{gpu}. This action space has two dimensions, corresponding to the power adjustments for the \gls*{cpu} and \gls*{gpu}, respectively. The range of the action space is [-1.0, 1.0], and then map it into power control. Within this interval, -1.0 signifies a decrease in power by the maximum amount, while 1.0 indicates an increase by the maximum amount. The description can be shown as follows:
\begin{itemize}
\item \gls*{cpu} Range: The interval [-1.0, 1.0] maps to [-2, 2], representing a maximum increase of 2 watts or a decrease of 2 watts.

\item \gls*{gpu} Range: The interval [-1.0, 1.0] maps to [-3, 3], representing a maximum increase of 3 watts or a decrease of 3 watts.
\end{itemize}

\subsection{Simulation Environment: YOLOv4 and NJTX2}

To evaluate the performance of our deep reinforcement learning-based \gls*{pmu-drl} framework, we set up a simulated environment for an energy-saving algorithm using the \gls*{njtx2} hardware platform. This platform is known for its low power consumption of 15 watts \cite{cui2019real} and is designed to run efficient state-of-the-art \gls*{ai} applications. The \gls*{njtx2} is a heterogeneous platform that includes the NVIDIA Jetpack \gls*{sdk} which provides libraries for \gls*{gpu}-accelerated computing, Linux drivers, and the Ubuntu operating system. The on-chip \gls*{gpu} can be programmed using NVIDIA's \gls*{cuda} \cite{paul2022simplified} for accelerating parallelisable algorithms and the \gls*{cudnn} library, which is optimised for \gls*{dnn}, is also installed to ensure maximum computational performance. Additionally, popular \gls*{ai} frameworks such as PyTorch and TensorFlow are integrated with the NVIDIA libraries to simplify the use of the \gls*{gpu} for \gls*{ai} developers. The \gls*{njtx2} is widely used in fields such as computer vision, data classification, deep learning and other areas that require intensive computing. It has a wide range of applications, including \gls*{iva}, drones, robots, gaming devices, \gls*{vr}, \gls*{ar} and portable medical devices.

The hardware specifications for the \gls*{njtx2} system, delineated in Table \ref{tab:TX2_info}, provide crucial information for configuring the simulation environment in this study. Demonstrating the computational requirements of large-scale deep neural network applications, the \gls*{yolo}v4 algorithm, detailed in \cite{bochkovskiy2020yolov4}, was employed. Furthermore, the \gls*{yolo}v4 algorithm stands out due to its C++ implementation, yielding lower latency compared to Python-based implementations of other \glspl*{cnn}. This independence from third-party libraries like PyTorch or TensorFlow streamlines \gls*{ai} computing. The C++ implementation of \gls*{yolo}v4 directly leverages the \gls*{cuda} instruction set, optimising \gls*{gpu} \gls*{ram} utilisation efficiency over Python implementations. Consequently, the algorithm is widely adopted in computing resource-constrained embedded environments. In contrast, Python, being a scripting language, requires an interpreter to interpret the code line by line, converting it into machine instructions each time the project is called, allowing the processor to execute the program. In contrast, C/C++ undergoes pre-compilation with gcc, resulting in a one-time compilation into a machine code file executed by the system. The approach enables multiple executions without compromising efficiency. The difference in underlying logic within the hardware system underscores that Python's execution efficiency is not on par with that of the C/C++ language. C/C++ language projects can also leverage the \gls*{cuda} library for accelerated computation, extending its utility beyond Python project development.

The devised hardware/software configuration effectively simulates the typical workloads imposed by cutting-edge \glspl*{cnn} running on edge devices. It authentically replicates human-like tasks, including detection, observation, recognition, and object identification—tasks commonly performed by individuals monitoring \gls*{cctv} systems. The application is executed on the hardware platform, capturing real-time power consumption data through the \gls*{sll} sensors. Subsequently, a simulation environment is constructed using this acquired information. Operating within this simulated setting, the \gls*{pmu-drl} framework learns and comprehends the hardware requirements for power consumption and performance of the application. Consequently, the \gls*{pmu-drl} framework optimises \gls*{gpu} scheduling to conserve power on the hardware platform.

In the \gls*{drl} framework, the assumption is made that the environment is Markov. Thereforem, the system's next state is presumed to depend solely on the current state and the action taken in the current state, excluding past states and actions. The \gls*{drl} model is trained and makes predictions based on this assumption. Nonetheless, recognition is given to the possibility that, in real-world heterogeneous computing environments, certain state variables or actions might have effects that persist into future time steps. Further exploration of this aspect is planned for future research.

\begin{table}[htbp]
\caption{\gls*{njtx2} specifications } \centering
\label{tab:TX2_info}
\begin{tabular}{p{75pt}p{130pt}}
\toprule 
Hardware Information & \gls*{njtx2} \\ 
\midrule 
\gls*{cpu} & Dual-Core for NVIDIA Denver 2 and Quad-Core for \gls*{arm} Cortex A57 \\
\gls*{gpu} & 256-core NVIDIA Pascal with \gls*{cuda} \\
Pipeline & 18-Stages, Out-of-Order, 3-way Issue \\
Cache & 48kB L1-I, 32kB L1-D, 2 MB L2 \\
Memory & 8 GB LPDDR4 \\
Operating System & Ubuntu 20.04 LTS\\ 
\bottomrule 
\end{tabular}
\end{table}

\begin{figure*}[htbp]
\centering
\includegraphics[width=0.99\textwidth]{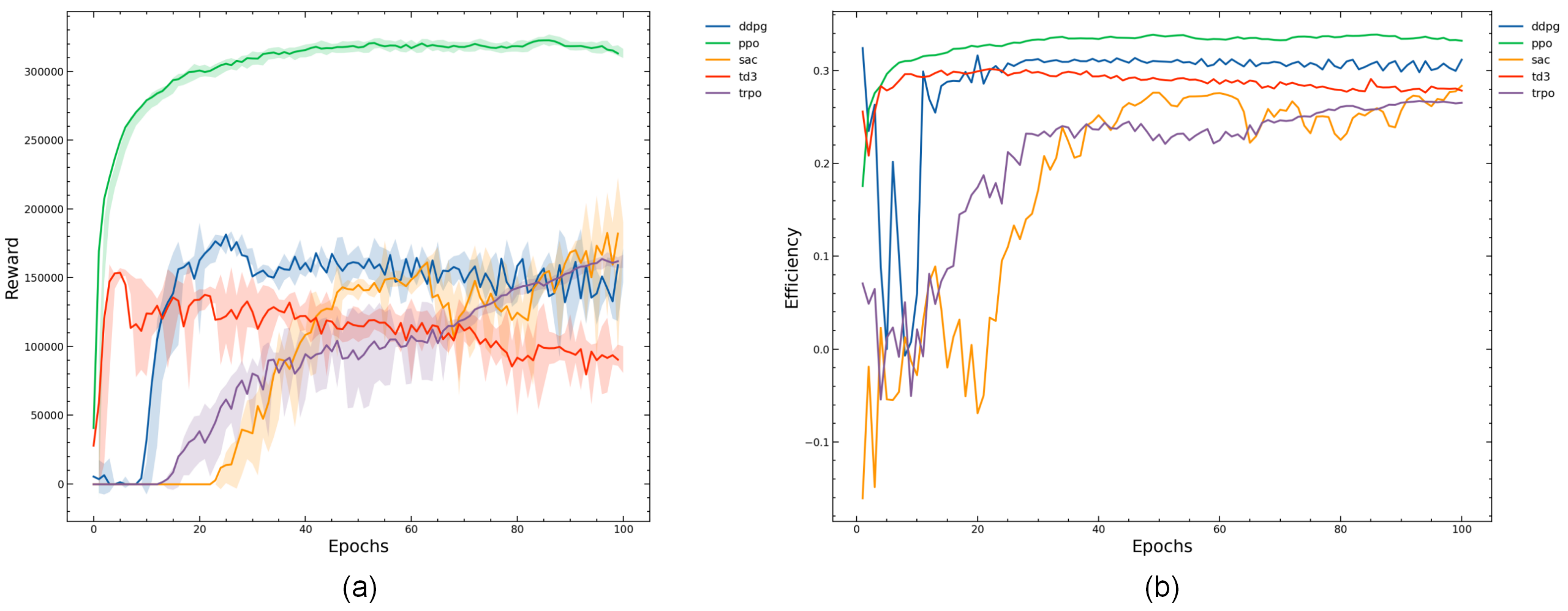}
\caption{Deep Reinforcement Learning Training Results.}
\label{fig:Training_Results}
\end{figure*}

\section{Experimental Evaluation} \label{sec:eval}

Figure \ref{fig:Training_Results} portrays the performance of the \gls*{drl} algorithm within the \gls*{pmu-drl} framework, revealing our method's primary learning and performance curves, and comparing with other popular deep reinforcement learning algorithms. All algorithms have been trained with the same epochs that are trained 100 epochs on each algorithm to make sure algorithms achieve stable learning output. Figure \ref{fig:Training_Results}.a provides insights into the agent's training process. It has shown the trend of the rewards‘ sum during algorithm learning. It is evident from the chart that the \gls*{ppo} algorithm exhibits the quickest increase in reward value during training and achieves the highest final reward compared to the other algorithms. Initially, all algorithms start with low rewards, but the \gls*{ppo} algorithm shows a sharp increase in rewards after approximately 5 epochs, rapidly surpassing the others. After around 25 epochs, the reward curve of \gls*{ppo} stabilizes, maintaining around 300,000, indicating good consistency and the highest performance. In contrast, the training reward curves of the other algorithms either increase more slowly (as with \gls*{trpo} and \gls*{td3}) or rise rapidly but result in lower final rewards (as with \gls*{ddpg} and \gls*{sac}). Not only does the \gls*{ppo} algorithm achieve the highest final reward, but its reward curve also exhibits smaller fluctuations, indicating better stability compared to the others.

For figure \ref{fig:Training_Results}.b, all algorithms experience fluctuations in efficiency, which is a common occurrence as the algorithms are still adapting to the environment. Over time of training, the \gls*{ppo} algorithm demonstrates higher and more stable efficiency. After an initial period of fluctuation, \gls*{ppo}'s efficiency quickly stabilizes and remains more than 0.34 (34\%), the highest among all algorithms. The other algorithms, such as \gls*{td3}, eventually reach a stable efficiency but at lower levels than \gls*{ppo}, and stabilising slower in the training process. \gls*{sac}, \gls*{ddpg} and \gls*{trpo} show lower efficiency for most of the training period, even exhibiting negative efficiency at certain points, indicating that the algorithms may misunderstand the power-saving experience rather than improve efficiency during those epochs. Overall, throughout the training period, \gls*{ppo} not only quickly achieves a high level of efficiency but also maintains it, indicating the best overall performance during the training process. This suggests that in a reinforcement learning framework, when efficiency is considered a key performance metric, the \gls*{ppo} algorithm may be the optimal choice.

The \gls*{drl} algorithm introduces a delay in assessing the impact of the current decision, necessitating additional time to ascertain if the intended outcome is achieved. Consequently, the adverse effects of decisions are not solely attributable to recent actions, but may also be influenced by errors that occurred in the previous period. The foundational principle of \gls*{drl} is to swiftly navigate away from the current state in the event of an error and judiciously adjust to facilitate convergence when executing correct actions.

In our experiment, power consumption is derived from sampled voltage and current data. Specifically, at each time point \textit{i}, denoted as (\textit{V[i]} and \textit{I[i]}), the voltage and current are captured by a sensor at a rate of 1 kHz. Consequently, the power at time point \textit{i}, \textit{P[i]}, is computed as the product of \textit{V[i]} and \textit{I[i]}. With a sampling rate \textit{f} of 1 kHz, the duration of each time point $\Delta$\textit{t} is \textit{1/f}. Accordingly, the energy consumption \textit{E}, representing the integral of power over time (as per Equation \ref{equation1}), can be approximated as the sum of the product of power and the duration at each time point: 
\begin{equation}\label{equation1}
E= \sum P\left [ i \right ]\ast \Delta t = \sum \left ( V\left [ i \right ] \ast I\left [ i \right ] \ast \left ( 1/f \right ) \right )
\end{equation}
For discrete time systems or simulations, this can be approximated as equation \label{equation2}:
\begin{equation}\label{equation2}
E= \sum P\left [ t \right ]\ast \Delta t
\end{equation}
where:
$\Delta$\textit{t} is the time step.
Given the context of the paper, the power consumption P(t) could be the power consumed by the \gls*{cpu} and \gls*{gpu} at a given time, which the proposed \gls*{pmu-drl} system is designed to minimise. The equation provides a precise method to quantify the energy consumption of the system, facilitating a fair and accurate comparison of the energy efficiency among different configurations and algorithms.

\begin{figure}[htbp]
\centering
\includegraphics[width=0.49\textwidth]{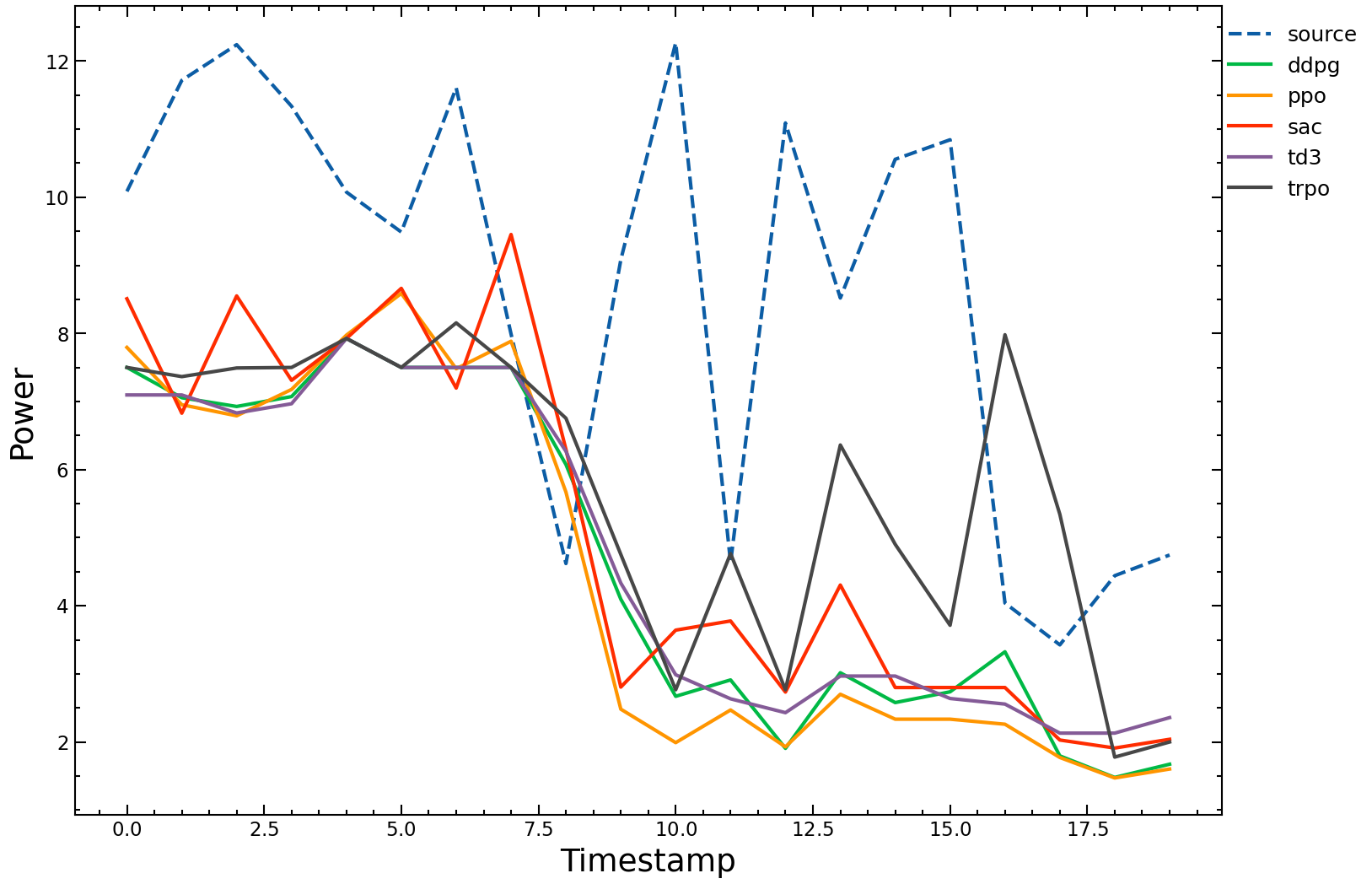}
\caption{Comparison of power consumption effects with and without \gls*{pmu-drl} framework.}
\label{fig:Power_Consumption}
\end{figure}

Figure \ref{fig:Power_Consumption} illustrates the learning output of the \gls*{pmu-drl} framework and other algorithms for the effect of energy-saving power consumption. The source curve shows the normal power consumption of the system, and it is compared with all implemented Deep Reinforcement Learning algorithms. Following the power curve trend, which produces our designed \gls*{pmu-drl} framework of the \gls*{ppo} algorithm is the best power consumption control effect. According to the simulation environment settings, the \gls*{pmu-drl} algorithm can improve the energy efficiency of the system by understanding the appropriate time for \gls*{gpu} context switching scheduling rules for different processing conditions required by hardware computing.

The \gls*{drl} agent is based on on-policy learning, which avoids the difference between the effect of the optimisation result and the theoretical value affecting the output of the decision. The simulation environment test reflects the different simulations of the running time of the popular program to achieve the overall reaction energy-saving effect. The ablation study of deep reinforcement algorithms for hardware power control resulting data, there are presented in Table \ref{tab:Complexity} and Table \ref{tab:energy}.

Table \ref{tab:Complexity} of algorithm complexity lists five deep reinforcement learning algorithms under the same network architecture that unifies layers and deep reinforcement algorithms setting, the compared results including two metrics: the number of floating-point operations (FLOPs) and the number of parameters (Parameters). Among these five algorithms, it can be seen that under the same deep reinforcement architecture, the \gls*{ppo} algorithm has the smallest computational complexity and requires the least number of parameters, which means it has the highest computational efficiency, the shortest training time, and the least storage space and memory resources required.

Table \ref{tab:energy} shows the power consumption and energy efficiency improvement comparison results for normal work mode and five deep reinforcement learning algorithms for hardware control over different time intervals (1s, 10s, 30s, 60s, and 90s). The influence of different time periods on energy consumption is examined through the continuous multi-object detection of the \gls*{yolo}v4 deep neural network algorithm in the test. All five deep reinforcement learning algorithms demonstrate significant energy savings compared to the normal work mode. The \gls*{drl} algorithm achieves the highest energy efficiency, reducing energy consumption by more than 34.6\% across all time intervals. Therefore, the algorithm can save power without compromising performance requirements, which depends on the algorithm to output more efficient scheduling rules for \gls*{gpu} context switching.

\begin{table}[htbp] 
\caption{Algorithms' Complexity} 
\label{tab:Complexity}
\centering
\begin{tabular}{p{120pt}p{45pt}p{45pt}p{45pt}p{50pt}p{50pt}}
\toprule 
\diagbox{Algs}{Energy}{Complexity} & FLOPs & Parameters \\ 
\midrule 
\gls*{drl}(PPO) &  9536  &  9797 \\ 
\gls*{ddpg} &  14336  &  14724  \\ 
\gls*{sac} &  24064  &  24712 \\ 
\gls*{td3} &  28672  &  29448  \\ 
\gls*{trpo} &  17600  &  17987 \\ 
\bottomrule 
\end{tabular}
\end{table}

\begin{table*}[htbp] 
\caption{Energy efficiency evaluation} 
\label{tab:energy}
\centering
\begin{tabular}{p{100pt}p{45pt}p{50pt}p{50pt}p{50pt}p{50pt}}
\toprule 
\diagbox{Algs}{\makecell{Energy \& \\ Efficiency}}{Times} & 1s & 10s & 30s & 60s & 90s \\ 
\midrule 
Normal Work Mode&  8.3623 mWh &  83.9054 mWh &  254.4317 mWh &  510.7967 mWh &  771.8151 mWh \\ 
\multirow{2}*{\gls*{drl} for Hardware Control} &  5.5356 mWh  &  54.9101 mWh &  166.6102 mWh &  333.7791 mWh &  504.4102 mWh \\ ~ & \textbf{( 33.8028\% )} & \textbf{( 34.5572\% )} & \textbf{( 34.5167\% )} & \textbf{( 34.6552\% )} & \textbf{( 34.6462\% )} \\ 
\multirow{2}*{\gls*{ddpg} for Hardware Control} &  5.5917 mWh &  55.2472 mWh &  167.6610 mWh &  336.0514 mWh &  508.2106 mWh \\ ~ & ( 33.1321\% ) & ( 34.1554\% ) & ( 34.1037\% ) & ( 34.2103\% ) & ( 34.1538\% ) \\ 
\multirow{2}*{\gls*{sac} for Hardware Control} &  5.9427 mWh &  58.8223 mWh &  177.9094 mWh &  356.5327 mWh &  537.8799 mWh \\ ~ & ( 28.9345\% ) & ( 29.8946\% ) & ( 30.0758\% ) & ( 30.2007\% ) & ( 30.3097\% ) \\ 
\multirow{2}*{\gls*{td3} for Hardware Control} &  5.6252 mWh &  55.7439 mWh &  169.1433 mWh &  339.0435 mWh &  512.4474 mWh \\ ~ & ( 32.7312\% ) & ( 33.5633\% ) & ( 33.5211\% ) & ( 33.6246\% ) & ( 33.6049\% ) \\ 
\multirow{2}*{\gls*{trpo} for Hardware Control} &  6.0435 mWh &  60.3145 mWh &  182.4971 mWh &  365.5448 mWh &  551.5803 mWh \\ ~ & ( 27.7291\% ) & ( 28.1161\% ) & ( 28.2727\% ) & ( 28.4363\% ) & ( 28.5347\% ) \\ 
\bottomrule 
\end{tabular}
\end{table*}

\begin{table*}[htbp]
\caption{Comparison energy efficiency with others work} 
\label{tab:Comparison_table}
\centering
\begin{tabular}{|p{40pt}|p{145pt}|p{183pt}|p{91pt}|}
\hline
Project & Hardware & Method & Energy Efficiency Improved \\ \hline
Our work & \gls*{njtx2} and \gls*{sll}  & Deep Reinforcement Learning algorithm to hardware control with \gls*{gpu} context switch rule & More than 34.6\%  average energy efficiency improved with the normal model \\ \hline
Yu et al. \cite{yu2020energy} & VCS-1 board based on Xilinx ZYNQ UltraScale+ \gls*{mpsoc} Chip and Sundance Lynsyn board & \gls*{drl} on the \gls*{mcu} to control and stop the clocks when no data are being exchanged via the I/Os & Up to 18\% power reduction compared with the original model \\ \hline
Saroj \emph{et. al} \cite{panda2022energy} & NVIDIA Jetson Nano & An application-deadline-aware data offloading scheme using deep reinforcement learning and \gls*{dvfs}  & The most energy reduction ranging between 9.68 and 10.35\% \\ \hline
Jose Nunez-Yanez \cite{nunez2018energy} & ZYNQ Z7020 and Zynq Ultrascale+ ZU9 & The extension and application of an adaptive voltage scaling framework called Elongate to a high-performance and reconfigurable binarised neural network & Energy efficiency between 5\%-23\% via a +/-1\% accuracy variation. \\ \hline
Sadrosadati \emph{et al} \cite{sadrosadati2019itap} & NVIDIA Tesla P100 & A idle-time-aware power management technique, which use finite state machine to effectively reduce the static energy consumption of \gls*{gpu} execution units by exploiting their idleness & Improves the static energy savings by an average of 16.9\% \\ \hline
\end{tabular}
\end{table*}

\section{Discussion} \label{sec:discussion}

The article introduces an energy-efficient framework tailored for edge \gls*{ai} hardware platforms, striking a balance between performance and power consumption. The framework simulates the real program's operational environment, showcasing the utilisation of the \gls*{lslb} sensors for capturing power consumption information. Hardware control is then facilitated through the \gls*{drl} algorithm, ultimately achieving energy-saving effects. The proposed approach diverges from conventional techniques like \gls*{dvfs} \cite{zhang2018double} and \gls*{avs} \cite{lacroix20196}, which necessitate data pre-processing of frequency and voltage information. Furthermore, compared to the \gls*{drl} algorithm employed in our earlier work \cite{yu2020energy}, the \gls*{pmu-drl} framework, employing self-adaptive feedback via the \gls*{ac} architecture, offers enhanced accuracy and stability. Operating solely on top-level system power consumption information, the \gls*{pmu-drl} framework orchestrates optimal energy-saving schemes for the entire system.

Table \ref{tab:Comparison_table} provides a comprehensive comparison of results obtained in this work with other similar projects from various sources, showcasing the superior efficiency and energy consumption performance of our proposed framework. The devised method builds upon a previously researched reinforcement learning architecture \cite{yu2020energy}, elevating it to a deep reinforcement learning algorithm. This upgrade not only enhances efficiency by over 16\% but also ensures a more stable output. In contrast, traditional methods relying on voltage and frequency control technologies, such as \gls*{dvfs} and \gls*{avs}, depend on data pre-processing for extensive data source analysis. Our method, in comparison, excels in achieving energy efficiency without compromising performance requirements and exhibits adaptability across different hardware platforms. The research underscores the adaptability of our proposed method across various platforms, evident in both theoretical considerations and algorithmic feasibility reflected in the test results. The deep reinforcement learning algorithm at the core of our adaptive architecture allows seamless matching with different hardware configurations without necessitating modifications to the decision-making environment. Furthermore, the algorithm's self-feedback adjustment through the \gls*{ac} architecture enables adaptive matching with unknown hardware parameters, effectively handling changes in hardware. It's essential to note that while a \gls*{drl} application optimises power consumption management in a heterogeneous computing environment, the energy consumption of the \gls*{drl} program itself was not measured in this research. The focus was primarily on leveraging the \gls*{drl} application to achieve optimal energy savings in a simulated environment, and future studies could explore and measure the energy consumption of the DRL program for a more comprehensive analysis.

\section{Conclusion} \label{sec:conclusion}
This research proposed an energy-efficient framework for edge \gls*{ai} hardware platforms that balances performance and power consumption through the use of a deep reinforcement learning algorithm. The results demonstrate that this method is highly effective, achieving a 34.6\% improvement in energy efficiency for sustainable system operation. Additionally, the proposed approach is based on self-adaption feedback by the \gls*{ac} architecture and does not require data pre-processing. This makes it a significant improvement over traditional methods such as \gls*{dvfs} and \gls*{avs}, which rely on data pre-processing of frequency and voltage information to achieve optimal performance. 

The results of this study provide strong theoretical support and algorithmic feasibility for extending this method to different platforms and hardware configurations in the future. One possible direction would be to investigate the scalability of the proposed method to larger and more complex edge \gls*{ai} systems. Additionally, further experimentation and testing could be conducted on different types of hardware platforms to validate the universality of the method. Another area for future research could be to explore the integration of other \gls*{ai} algorithms, such as neural networks or evolutionary algorithms, to enhance the performance and energy efficiency of the proposed method. Finally, it would be desirable to apply the proposed method to other types of systems, such as data centres or cloud computing environments, to further reduce energy consumption and without detorating the computational performance.

\section{Acknowledgments}
This research project has been made possible with the support of Sundance Multiprocessor Technology Ltd, through the European Union's Horizon 2020 research and innovation program projects: Optimisation for Energy Consumption and Performance Trade-off, under grant agreement No. 779656 and VineScout, under grant agreement No. 737669. The hardware support power monitor system used in this research was provided by Sundance Multiprocessor Technology Ltd. We would like to express our gratitude for their invaluable assistance and support. (Figure \ref{fig:sll} illustrates the hardware support provided by Sundance Multiprocessor Technology Ltd.). This work is supported by EPSRC projects, CHEDDAR EP/X040518/1 and CHEDDAR Uplift EP/Y037421/1.

\bibliographystyle{IEEEtran}
\bibliography{IEEE_TMC}

\begin{thebibliography}{10}
\providecommand{\url}[1]{#1}
\csname url@samestyle\endcsname
\providecommand{\newblock}{\relax}
\providecommand{\bibinfo}[2]{#2}
\providecommand{\BIBentrySTDinterwordspacing}{\spaceskip=0pt\relax}
\providecommand{\BIBentryALTinterwordstretchfactor}{4}
\providecommand{\BIBentryALTinterwordspacing}{\spaceskip=\fontdimen2\font plus
\BIBentryALTinterwordstretchfactor\fontdimen3\font minus
  \fontdimen4\font\relax}
\providecommand{\BIBforeignlanguage}[2]{{%
\expandafter\ifx\csname l@#1\endcsname\relax
\typeout{** WARNING: IEEEtran.bst: No hyphenation pattern has been}%
\typeout{** loaded for the language `#1'. Using the pattern for}%
\typeout{** the default language instead.}%
\else
\language=\csname l@#1\endcsname
\fi
#2}}
\providecommand{\BIBdecl}{\relax}
\BIBdecl

\bibitem{zhang2018double}
Q.~Zhang, M.~Lin, L.~T. Yang, Z.~Chen, S.~U. Khan, and P.~Li, ``A double deep
  q-learning model for energy-efficient edge scheduling,'' \emph{IEEE
  Transactions on Services Computing}, vol.~12, no.~5, pp. 739--749, 2018.

\bibitem{yu2020energy}
Z.~Yu, P.~Machado, A.~Zahid, A.~M. Abdulghani, K.~Dashtipour, H.~Heidari, M.~A.
  Imran, and Q.~H. Abbasi, ``Energy and performance trade-off optimization in
  heterogeneous computing via reinforcement learning,'' \emph{Electronics},
  vol.~9, no.~11, p. 1812, 2020.

\bibitem{kumar2020conservative}
A.~Kumar, A.~Zhou, G.~Tucker, and S.~Levine, ``Conservative q-learning for
  offline reinforcement learning,'' \emph{Advances in Neural Information
  Processing Systems}, vol.~33, pp. 1179--1191, 2020.

\bibitem{yu2022radar}
Z.~Yu, A.~Taha, W.~Taylor, A.~Zahid, K.~Rajab, H.~Heidari, M.~A. Imran, and
  Q.~H. Abbasi, ``A radar-based human activity recognition using a novel 3d
  point cloud classifier,'' \emph{IEEE Sensors Journal}, 2022.

\bibitem{9878152}
Z.~Yu, A.~Zahid, A.~Taha, W.~Taylor, J.~L. Kernec, H.~Heidari, M.~A. Imran, and
  Q.~H. Abbasi, ``An intelligent implementation of multi-sensing data fusion
  with neuromorphic computing for human activity recognition,'' \emph{IEEE
  Internet of Things Journal}, pp. 1--1, 2022.

\bibitem{wang2022deep}
X.~Wang, S.~Wang, X.~Liang, D.~Zhao, J.~Huang, X.~Xu, B.~Dai, and Q.~Miao,
  ``Deep reinforcement learning: a survey,'' \emph{IEEE Transactions on Neural
  Networks and Learning Systems}, 2022.

\bibitem{wang2022energy}
Q.~Wang, X.~Mei, H.~Liu, Y.-W. Leung, Z.~Li, and X.~Chu, ``Energy-aware
  non-preemptive task scheduling with deadline constraint in dvfs-enabled
  heterogeneous clusters,'' \emph{IEEE Transactions on Parallel and Distributed
  Systems}, vol.~33, no.~12, pp. 4083--4099, 2022.

\bibitem{nabavinejad2022coordinated}
S.~M. Nabavinejad, S.~Reda, and M.~Ebrahimi, ``Coordinated batching and dvfs
  for dnn inference on gpu accelerators,'' \emph{IEEE transactions on parallel
  and distributed systems}, vol.~33, no.~10, pp. 2496--2508, 2022.

\bibitem{zamani2023greenmd}
H.~Zamani, L.~Bhuyan, J.~Chen, and Z.~Chen, ``Greenmd: Energy-efficient matrix
  decomposition on heterogeneous multi-gpu systems,'' \emph{ACM Transactions on
  Parallel Computing}, vol.~10, no.~2, pp. 1--23, 2023.

\bibitem{wulf2020low}
C.~Wulf, M.~Willig, and D.~G{\"o}hringer, ``Low power scheduling of periodic
  hardware tasks in flash-based fpgas,'' in \emph{2020 IEEE Nordic Circuits and
  Systems Conference (NorCAS)}.\hskip 1em plus 0.5em minus 0.4em\relax IEEE,
  2020, pp. 1--7.

\bibitem{wulf2022rtos}
C.~Wulf, M.~Willig, and D.~Goehringer, ``Rtos-supported low power scheduling of
  periodic hardware tasks in flash-based fpgas,'' \emph{Microprocessors and
  Microsystems}, p. 104566, 2022.

\bibitem{wilson2021embedded}
A.~Wilson, A.~Kumar, A.~Jha, and L.~R. Cenkeramaddi, ``Embedded sensors,
  communication technologies, computing platforms and machine learning for
  uavs: A review,'' \emph{IEEE Sensors Journal}, vol.~22, no.~3, pp.
  1807--1826, 2021.

\bibitem{elster2022nvidia}
A.~C. Elster and T.~A. Haugdahl, ``Nvidia hopper gpu and grace cpu
  highlights,'' \emph{Computing in Science \& Engineering}, vol.~24, no.~2, pp.
  95--100, 2022.

\bibitem{bochkovskiy2020yolov4}
A.~Bochkovskiy, C.-Y. Wang, and H.-Y.~M. Liao, ``Yolov4: Optimal speed and
  accuracy of object detection,'' \emph{arXiv preprint arXiv:2004.10934}, 2020.

\bibitem{redmon2016you}
J.~Redmon, S.~Divvala, R.~Girshick, and A.~Farhadi, ``You only look once:
  Unified, real-time object detection,'' in \emph{Proceedings of the IEEE
  conference on computer vision and pattern recognition}, 2016, pp. 779--788.

\bibitem{legaspi2021detection}
K.~R.~B. Legaspi, N.~W.~S. Sison, and J.~F. Villaverde, ``Detection and
  classification of whiteflies and fruit flies using yolo,'' in \emph{2021 13th
  International Conference on Computer and Automation Engineering
  (ICCAE)}.\hskip 1em plus 0.5em minus 0.4em\relax IEEE, 2021, pp. 1--4.

\bibitem{santoni2022traveling}
F.~Santoni, A.~De~Angelis, A.~Moschitta, and P.~Carbone, ``A traveling standard
  for calibration of battery impedance measurement systems,'' \emph{IEEE
  Transactions on Instrumentation and Measurement}, vol.~71, pp. 1--8, 2022.

\bibitem{djupdal2021lynsyn}
A.~Djupdal, B.~Gottschall, F.~Ghasemi, and M.~Jahre, ``Lynsyn and lynsynlite:
  The sthem power measurement units,'' in \emph{Towards Ubiquitous Low-power
  Image Processing Platforms}.\hskip 1em plus 0.5em minus 0.4em\relax Springer,
  2021, pp. 93--114.

\bibitem{callebaut2021art}
G.~Callebaut, G.~Leenders, J.~Van~Mulders, G.~Ottoy, L.~De~Strycker, and
  L.~Van~der Perre, ``The art of designing remote iot devices—technologies
  and strategies for a long battery life,'' \emph{Sensors}, vol.~21, no.~3, p.
  913, 2021.

\bibitem{sabogal2019recon}
S.~Sabogal, A.~George, and G.~Crum, ``Recon: A reconfigurable cnn acceleration
  framework for hybrid semantic segmentation on hybrid socs for space
  applications,'' in \emph{2019 IEEE Space Computing Conference (SCC)}.\hskip
  1em plus 0.5em minus 0.4em\relax IEEE, 2019, pp. 41--52.

\bibitem{ye2023real}
T.~Ye, W.~Qin, Z.~Zhao, X.~Gao, X.~Deng, and Y.~Ouyang, ``Real-time object
  detection network in uav-vision based on cnn and transformer,'' \emph{IEEE
  Transactions on Instrumentation and Measurement}, vol.~72, pp. 1--13, 2023.

\bibitem{brunacci2023fusion}
V.~Brunacci and A.~De~Angelis, ``Fusion of uwb and magnetic ranging systems for
  robust positioning,'' \emph{IEEE Transactions on Instrumentation and
  Measurement}, 2023.

\bibitem{oliveira2019interactive}
B.~G. Oliveira and J.~Lobo, ``Interactive demonstration of an energy efficient
  yolov3 implementation in reconfigurable logic,'' in \emph{2019 5th Experiment
  International Conference (exp. at'19)}.\hskip 1em plus 0.5em minus
  0.4em\relax IEEE, 2019, pp. 235--236.

\bibitem{khaing2018development}
Z.~M. Khaing, Y.~Naung, and P.~H. Htut, ``Development of control system for
  fruit classification based on convolutional neural network,'' in \emph{2018
  IEEE conference of russian young researchers in electrical and electronic
  engineering (EIConRus)}.\hskip 1em plus 0.5em minus 0.4em\relax IEEE, 2018,
  pp. 1805--1807.

\bibitem{yu2023robust}
L.~Yu, E.~Yang, B.~Yang, Z.~Fei, and C.~Niu, ``A robust learned feature-based
  visual odometry system for uav pose estimation in challenging indoor
  environments,'' \emph{IEEE Transactions on Instrumentation and Measurement},
  2023.

\bibitem{tino2020simt}
A.~Tino, C.~Collange, and A.~Seznec, ``Simt-x: Extending single-instruction
  multi-threading to out-of-order cores,'' \emph{ACM Transactions on
  Architecture and Code Optimization (TACO)}, vol.~17, no.~2, pp. 1--23, 2020.

\bibitem{schulman2017proximal}
J.~Schulman, F.~Wolski, P.~Dhariwal, A.~Radford, and O.~Klimov, ``Proximal
  policy optimization algorithms,'' \emph{arXiv preprint arXiv:1707.06347},
  2017.

\bibitem{huang2020deep}
B.~Huang and J.~Wang, ``Deep-reinforcement-learning-based capacity scheduling
  for pv-battery storage system,'' \emph{IEEE Transactions on Smart Grid},
  vol.~12, no.~3, pp. 2272--2283, 2020.

\bibitem{su2021value}
J.~Su, S.~Adams, and P.~Beling, ``Value-decomposition multi-agent
  actor-critics,'' in \emph{Proceedings of the AAAI conference on artificial
  intelligence}, vol.~35, no.~13, 2021, pp. 11\,352--11\,360.

\bibitem{cui2019real}
H.~Cui and N.~Dahnoun, ``Real-time stereo vision implementation on nvidia
  jetson tx2,'' in \emph{2019 8th Mediterranean Conference on Embedded
  Computing (MECO)}.\hskip 1em plus 0.5em minus 0.4em\relax IEEE, 2019, pp.
  1--5.

\bibitem{paul2022simplified}
S.~Paul, S.~Mulani, N.~Daimary, and M.~S. Singh,
  ``Simplified-delay-multiply-and-sum-based promising beamformer for real-time
  photoacoustic imaging,'' \emph{IEEE Transactions on Instrumentation and
  Measurement}, vol.~71, pp. 1--9, 2022.

\bibitem{panda2022energy}
S.~K. Panda, M.~Lin, and T.~Zhou, ``Energy efficient computation offloading
  with dvfs using deep reinforcement learning for time-critical iot
  applications in edge computing,'' \emph{IEEE Internet of Things Journal},
  2022.

\bibitem{nunez2018energy}
J.~Nunez-Yanez, ``Energy proportional neural network inference with adaptive
  voltage and frequency scaling,'' \emph{IEEE Transactions on Computers},
  vol.~68, no.~5, pp. 676--687, 2018.

\bibitem{sadrosadati2019itap}
M.~Sadrosadati, S.~B. Ehsani, H.~Falahati, R.~Ausavarungnirun, A.~Tavakkol,
  M.~Abaee, L.~Orosa, Y.~Wang, H.~Sarbazi-Azad, and O.~Mutlu, ``Itap:
  Idle-time-aware power management for gpu execution units,'' \emph{ACM
  Transactions on Architecture and Code Optimization (TACO)}, vol.~16, no.~1,
  pp. 1--26, 2019.

\bibitem{lacroix20196}
M.-A. LaCroix, H.~Wong, Y.~H. Liu, H.~Ho, S.~Lebedev, P.~Krotnev, D.~A.
  Nicolescu, D.~Petrov, C.~Carvalho, S.~Alie \emph{et~al.}, ``6.2 a 60gb/s
  pam-4 adc-dsp transceiver in 7nm cmos with snr-based adaptive power scaling
  achieving 6.9 pj/b at 32db loss,'' in \emph{2019 IEEE International
  Solid-State Circuits Conference-(ISSCC)}.\hskip 1em plus 0.5em minus
  0.4em\relax IEEE, 2019, pp. 114--116.

\end{thebibliography}

\begin{IEEEbiography}[{\includegraphics[width=1in,height=1.25in,clip,keepaspectratio]{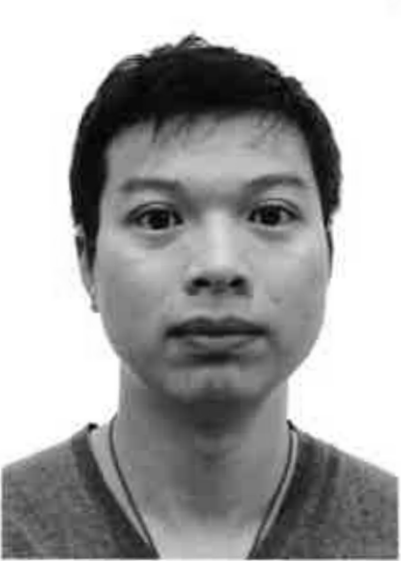}}]{Zheqi Yu}
received a Bachelor degree in electronic information engineering in 2014, and a Master degree in Information Technology in 2015 from the University of Wolverhampton. He received his Ph.D. degree in Electronic and Electrical engineering from University of Glasgow, U.K., in 2022. Currently, he is Research Engineer at Opteran Ltd, UK, which works on Neuromorphic Hardware design. His other research interests include Neuromorphic computing, hardware design, embedded systems, signal processing, power efficiency, and programmable devices.
\end{IEEEbiography}

\begin{IEEEbiography}[{\includegraphics[width=1in,height=1.25in,clip,keepaspectratio]{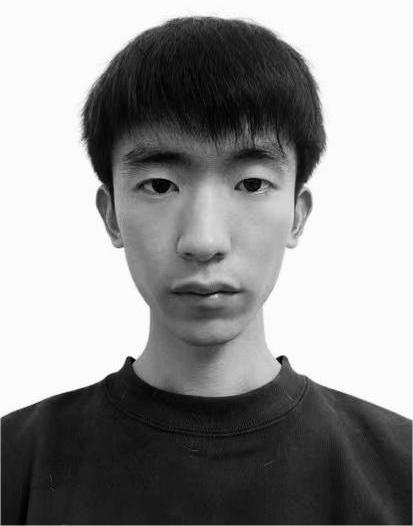}}]{Chao Zhang} 
received a Bachelor's degree in Computer Science and Technology in 2019 from the Shanghai Institute of Technology. He is currently working as a research engineer at LAN-XEN, Technology, INC., where his work is focused on research and software development in the area of artificial intelligence penetration testing. In addition to his primary job, he is also deeply passionate about researching the fields of cybersecurity and reinforcement learning.

\end{IEEEbiography}

\begin{IEEEbiography}[{\includegraphics[width=1in,height=1.25in,clip,keepaspectratio]{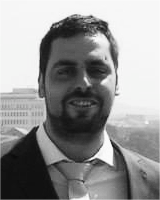}}]{Pedro Machado}
received his MSc in Electrical and Computers Engineering from the University of Coimbra (2012) and his PhD in Computer Science from Nottingham Trent University (2022). Dr Machado is a Senior Lecturer in Computer Science, Course Leader for MSc Artificial Intelligence and the secretary for the IEEE Systemic Innovation Special Interest Group (SISIG)
Dr Machado’s research interests includes neuromorphic engineering, edge computer vision, bio-inspired computing, robotics and intelligent sensors, retinal cell understanding, biological nervous system modelling, spiking neural networks, robotics and autonomous systems, and neuromorphic hardware, aquaculture, endangered/invasive underwater species.
\end{IEEEbiography}

\begin{IEEEbiography}[{\includegraphics[width=1in,height=1.25in,clip,keepaspectratio]{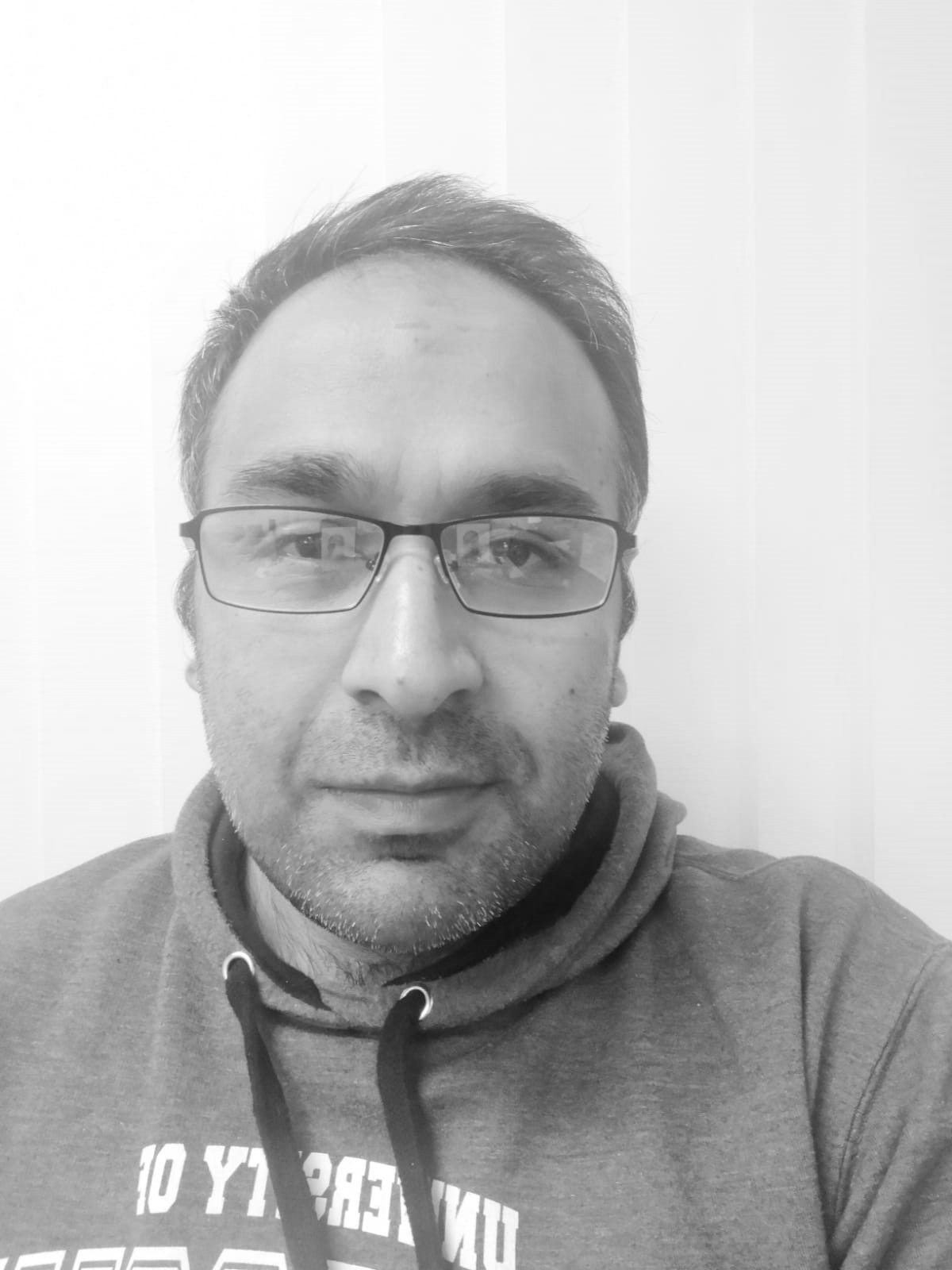}}]{Adnan Zahid}
received the B.Sc. (Hons.) degree in electronics and communications engineering from Glasgow Caledonian University and the M.Sc. degree in electronics and electrical engineering from the University of Strathclyde in 2016. He is currently pursuing the Ph.D. Research degree with the University of Glasgow. His current research interests encompasses machine learning to monitor plant’s health for precision agriculture applications, and detection of water stress in leaves by integrating deep learning and terahertz sensing.
\end{IEEEbiography}

\begin{IEEEbiography}[{\includegraphics[width=1in,height=1.25in,clip,keepaspectratio]{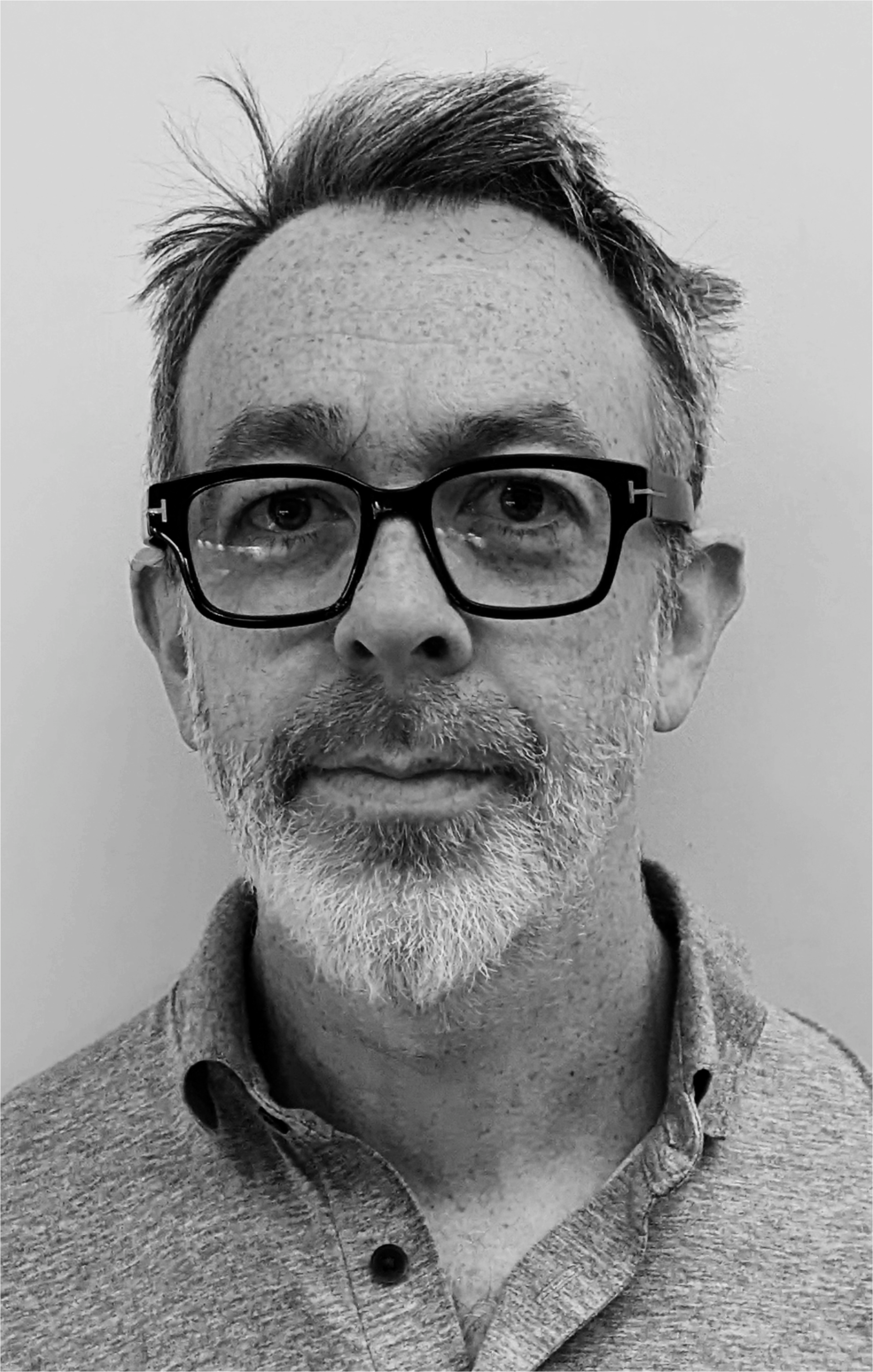}}]{Tim Fernandez-Hart} 
has a B.Sc. (Hons) in Genetics, an M.Sc. in Industrial Biotechnolgy, a M.Sc. in Medical Ultrasound and received a B.Sc. with first-class honours in Mathematics and Statistics from The Open University, UK in 2021. He is currently working towards a Ph.D. in Computer Systems Research at Brunel University, London, UK. His research interests include AI, neuromorphic engineering, spiking neural networks, event-based sensors and computer arithmetic. 

\end{IEEEbiography}

\begin{IEEEbiography}[{\includegraphics[width=1in,height=1.25in,clip,keepaspectratio]{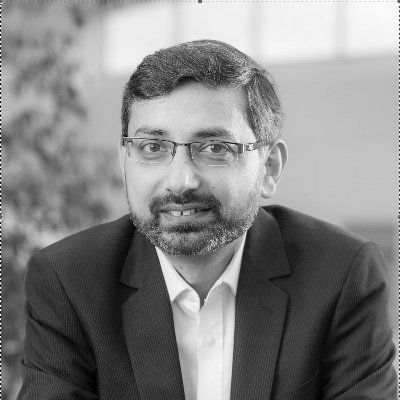}}]{Muhammad Imran}
(IEEE Fellow 2023) received his M.Sc. (Distinction) and Ph.D. degrees from Imperial College London, UK, in 2002 and 2007, respectively. He is a Professor in Communication Systems in the University of Glasgow, Dean University of Glasgow UESTC, Head of Autonomous Systems and Connectivity research division, Head of Communications Sensing and Imaging (CSI) research group and Director of Glasgow UESTC Centre of Educational Development and Innovation. He also serves as an affiliate Professor at the University of Oklahoma, USA; Adjunct Research Professor AIRC, Ajman University UAE and a visiting Professor at 5G Innovation centre, University of Surrey, UK. He has led a number of multimillion-funded international research projects and the "new physical layer" work area for 5G innovation centre at Surrey. He is a Senior Fellow of the Higher Education Academy, U.K. Prof. Imran is a Fellow of IEEE.
\end{IEEEbiography}

\begin{IEEEbiography}[{\includegraphics[width=1in,height=1.25in,clip,keepaspectratio]{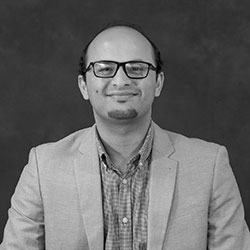}}]{Qammer Abbasi}
(SM`16) received his BSc and MSc degree in electronics and telecommunication engineering from University of Engineering and Technology (UET), Lahore, Pakistan (with distinction). He received his Ph.D. degree in Electronic and Electrical engineering from Queen Mary University of London (QMUL), U.K., in Jan., 2012. Currently, he is Professor of Applied Electromagnetics \& Sensing and theme lead for connecting people with the James Watt School of Engineering, University of Glasgow, U.K., deputy head for Communication Sensing and Imaging group, past Program Director for Dual PhD Degree, deputy theme lead for Quantum in the University’s Advance Research Centre, Co-Manager for RF and terahertz laboratory, lead for healthcare and Internet of things use cases with 5G Center Urban testbed and Project Manager for EON XR Centre. He has grant portfolio of £9M. 
\end{IEEEbiography}

\end{document}